\documentclass[10pt,a4paper,english]{revtex4}

\usepackage[T1]{fontenc}
\usepackage[latin1]{inputenc}
\usepackage{pifont}
\usepackage{graphicx}
\usepackage{amssymb}

\makeatletter
\AtBeginDocument{

}

\makeatother

\usepackage{babel}

\begin{document}

\title{Rippling of Graphene}

\author{Rebecca C. Thompson-Flagg, Maria J. B. Moura, and M. Marder}

\affiliation{Center for Nonlinear Dynamics and Department of Physics, The University
of Texas at Austin, Austin TX, USA}

\maketitle
\textbf{Meyer}\textbf{\emph{ et al}}\textbf{\cite{Meyer2007} found
that free-standing graphene sheets, just one atom thick, display spontaneous
ripples. The ripples are of order 2-20 $\mbox{\AA}$ high and 20-200
$\mbox{\AA}$ wide. The sheets in which they appear are only one atom
thick, and extend for around 5000 $\mbox{\AA}$ through vacuum between
metal struts that support them. Other groups have since created free-standing
graphene as well\cite{Garcia-Sanchez2008,Bolotin2008,Bolotin2008a},
and similar ripples have been found for graphene on a glass substrate\cite{Geringer2008}.
Here we show that these ripples can be explained as a consequence
of adsorbed molecules sitting on random sites. The adsorbates cause
the bonds between carbon atoms to lengthen slightly. Static buckles
then result from a mechanism like the one that leads to buckling of
leaves; buckles caused by roughly 20\% coverage of adsorbates are
consistent with experimental observations. We explain why this mechanism
is more likely to explain rippled than thermal fluctuations or the
Mermin-Wagner theorem, which have previously been invoked.}

Fasolino, Los, and Katsnelson\cite{Fasolino2007} and Abdepour \emph{et
al. }\cite{Abedpour2007} studied the possibility that the ripples
are due to thermal fluctuations. It seems unlikely to us that this
mechanism explains the observations. Fasolino \emph{et al.} report
that the bending stiffness $\kappa$ of graphene is on the order of
1 eV. According to Abraham and Nelson\cite{Abraham1990}, the root-mean-square
fluctuations of a thin sheet are \begin{equation}
\langle h^{2}\rangle\approx\frac{k_{B}T}{2\pi}\int_{L^{-1}}^{a^{-1}}\frac{qdq}{q^{4}\kappa_{R}(q)}\approx\frac{k_{B}T}{4\pi\kappa}L^{2},\label{eq:h2}\end{equation}
where $T$ is the temperature (the experiments are performed at room
temperature), and $L\sim$5000 $\mbox{\AA}$  is the length of the
sheet. For the purpose of making order of magnitude estimates, we
replace $\kappa_{R}(q)$ by the constant $\kappa$ , and get a value
$h\approx$200 $\mbox{\AA}$, which is 10 to 100 times larger than
the experimental observations. A more vexing problem is created by
the natural oscillation frequencies of the ripples. The kinetic and
bending energy of a thin sheet are approximately \cite{Landau.86}
\begin{equation}
E=\int d^{2}r\:\left[\frac{1}{2}\rho\dot{h}^{2}+\frac{\kappa}{2}\left(\nabla^{2}h\right)^{2}\right],\label{eq:energy}\end{equation}
where $\rho$ is the mass per area. Thus thermally excited ripples
of wavenumber $k$ should be oscillating at a frequency $\omega$
given by\begin{equation}
\omega^{2}=\frac{\kappa}{\rho}k^{4}.\label{eq:w2}\end{equation}
 For waves of scale 200 $\mbox{\AA}$ , the oscillation frequency
comes out to be of order $10^{10}$ Hz, while for waves of 50 $\mbox{\AA}$
the frequency is higher, $10^{11}$ Hz. These time scales are slow
compared with electron dynamics, and thermal oscillations of graphene
could affect transport properties\cite{Fasolino2007,Abedpour2007}.
However diffraction through such a rapidly oscillating membrane should
be sensitive only to the time average of the density-density correlation
function, and that should be completely periodic on time scales much
longer than $10^{-10}$ s, producing sharp Bragg peaks. However, the
analysis of Meyer \emph{et al} obtains agreement with the broadened
peaks they observed by assuming that the graphene has static ripples.
Thus the assumption that the observed wrinkling in the graphene layers
is purely due to thermal fluctuations seems unlikely.

Another explanation of the ripples\cite{Meyer2007,Geringer2008}
invokes the Mermin-Wagner theorem\cite{Mermin.66a,Mermin.68}, according
to which a truly two-dimensional crystal does not exist except at
zero temperature. We do not think that the Mermin-Wagner theorem is
actually significant here. What this theorem says is that when a two-dimensional
crystal remains completely planar, it will undergo rotations at large
distances that cause a breakdown of long-range order. In the case
of graphene at room temperature, the scale $l$ on which this happens
can be estimated from the expression

\begin{equation}
l\sim a\exp[Ga^{2}/k_{B}T],\label{eq:MW}\end{equation}
where $a$ is a lattice spacing and $G$ is the shear modulus of graphene.
Since the bulk shear modulus $\mu\approx G/a$ of graphite is 440
GPa\cite{Popov2000}, and the lattice spacing is greater than $a\gtrsim1$\AA,
one has $l>10^{30}$ m. So angular rotation of the crystal on the
scale of microns should be irrelevant. 

Thus, having decided that neither thermal fluctuations nor the Mermin-Wagner
theorem explain rippled graphene well, we decided to search for a
mechanism that could produce ripples in the ground state.

\begin{figure}
\begin{centering}
\includegraphics[scale=0.8]{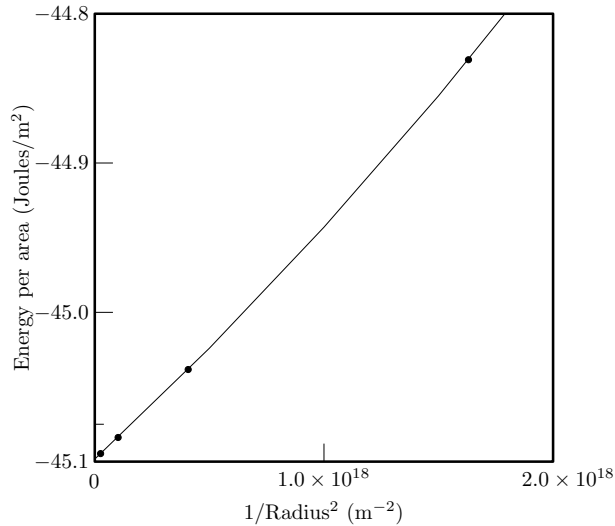}
\par\end{centering}

\caption{Energy per area of graphene cylinders plotted versus $1/R^{2}$ where
$R$ is the radius of the cylinder. Using $(1\ \ 0)$ and $(1/2\ \ \sqrt{3/4})$
as primitive vectors, the honeycomb lattice was wrapped around the
$\hat{y}$ axis. Periodic boundary conditions are imposed along the
axis direction. \label{fig:Energy-per-area} }

\end{figure}

Experience with mechanics of thin sheets led us to wonder if graphene
might be rippling in a fashion similar to leaves and torn plastic\cite{Sharon.02,Sharon.AmSci.04,Marder.PTD.07}.
In leaves and torn plastic, buckling results from a change in metric.
For graphene, this mechanism can lead to buckling if for some reason
the equilibrium distance between carbon atoms changes at some but
not all parts of the sheet. To investigate this idea, we used the
Modified Embedded Atom Method (MEAM)\cite{Baskes.92,Baskes.94} potential,
with parameters of Lee and Lee\cite{Lee2005}. To find the bending
stiffness, we constructed graphene cylinders of various radii $R$
and plotted energy per area versus $1/R^{2},$ as shown in Figure
\ref{fig:Energy-per-area}. The energy per area of a thin elastic
cylinder is $E=\frac{1}{2}\kappa/R^{2}$ where $\kappa$ is a bending
modulus. We extract a modulus of $\kappa=$1.77eV from these data,
which is higher than the value of 1.17eV employed by Fasolino \emph{et.
al.\cite{Fasolino2007} }although of the same order of magnitude.
To the extent the MEAM potential overstates the bending energy of
graphene, it will tend to underestimate the scale of ripples.

\begin{figure}
\begin{centering}
\includegraphics[width=6in]{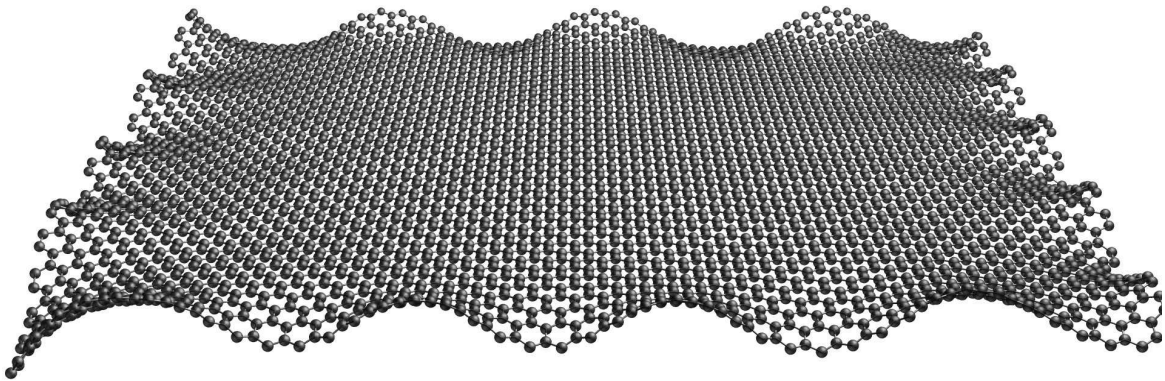}
\par\end{centering}

\caption{Ripples in graphene produced by edge effects alone in graphene sheet
simulated by MEAM, 100$\times$ 100 \AA.\label{fig:Ripples-in-graphene}}

\end{figure}

A first possibility we checked is that under-coordinated carbon atoms
at the edge of the sheet might create ripples at the edge that propagate
all the way in to the interior of the samples. Figure \ref{fig:Ripples-in-graphene}
shows a system $100\ \mbox{\AA}\times100\ \mbox{\AA}$ in size. The
edges are free, and the system is initially entirely flat, except
that atoms are randomly displaced from initial conditions by around
$10^{-2}$ $\mbox{\AA}$. Then the system is allowed to relax toward
a minimum energy state through damped dynamics. 

\begin{figure}
\begin{centering}
\includegraphics{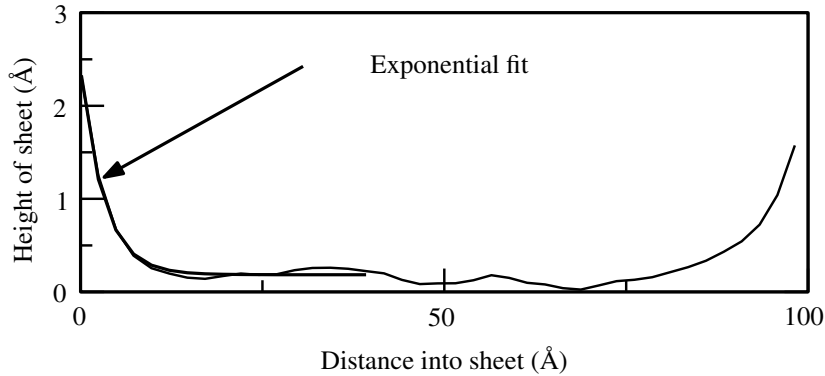}
\par\end{centering}

\caption{Slice through system shown in Figure \ref{fig:Ripples-in-graphene}
showing that the ripples decay with a characteristic distance of about
3 $\mbox{\AA}$ away from the edge of the graphene sheet.\label{fig:Slice-through-system}}

\end{figure}

Edge effects do indeed create ripples, and they are of the right size,
around 30 $\mbox{\AA}$ in width and $\mbox{10 \AA}$ in height. However,
the ripples occur only at the edge of the system, and the amplitude
decays to zero on the scale of 3.2 $\mbox{\AA}$ (Figure \ref{fig:Slice-through-system}).
Thus edge effects alone cannot explain the presence of ripples throughout
a sample of size 5000$\mbox{\AA}$.

In correspondence with A. Geim, we learned that  OH molecules could
be expected to be adsorbed on the surface of the graphene sheet. Other
molecules such as water and hydrogen may also be present\cite{Echtermeyer2007,Haluka,Moser2008}.
Details of the molecule are not likely to be critical, and we discuss
OH in order to have a specific example. The density of the adsorbates
is not known, so we have treated it as a free parameter, and considered
the effects of randomly placing OH molecules on graphene. According
to Xu \emph{et al}\cite{Xu2007}, attaching an OH molecule to a carbon
surface has the consequence of increasing the length of two adjoining
C-C bonds by around 10\% (see their Figure 2, L0D+OH). Rather than
directly simulating the interactions of graphene with OH, we simply
chose carbon atoms randomly from the lattice, and increased the equilibrium
length to two randomly chosen neighbors by this amount. This corresponds
to a small-scale change in the metric of the sheet. For the bonds
we wished to stretch, the MEAM parameter $R_{c}$ was increased to
1.48 $\mbox{\AA}$ from $1.42\ \mbox{\AA}$. This value was chosen
because if used for the entire crystal, it produces an equilibrium
lattice parameter close to what Xu\emph{ et al} find for the C-C bonds
stretched by OH. The energy associated with the OH attachment is .17
eV.

\begin{figure}
\begin{centering}
\includegraphics[width=6in]{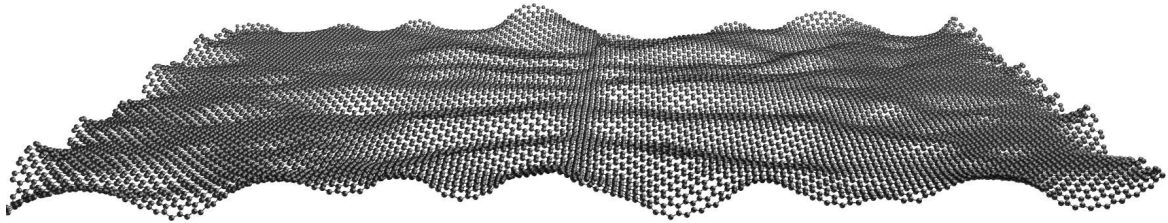}
\par\end{centering}

\caption{Ripples in graphene produced by 20\% coverage of OH, 200$\times$200
$\mbox{\AA}$ system. \label{fig:Ripples-in-graphene2}.}

\end{figure}

\begin{figure}
\begin{centering}
\includegraphics{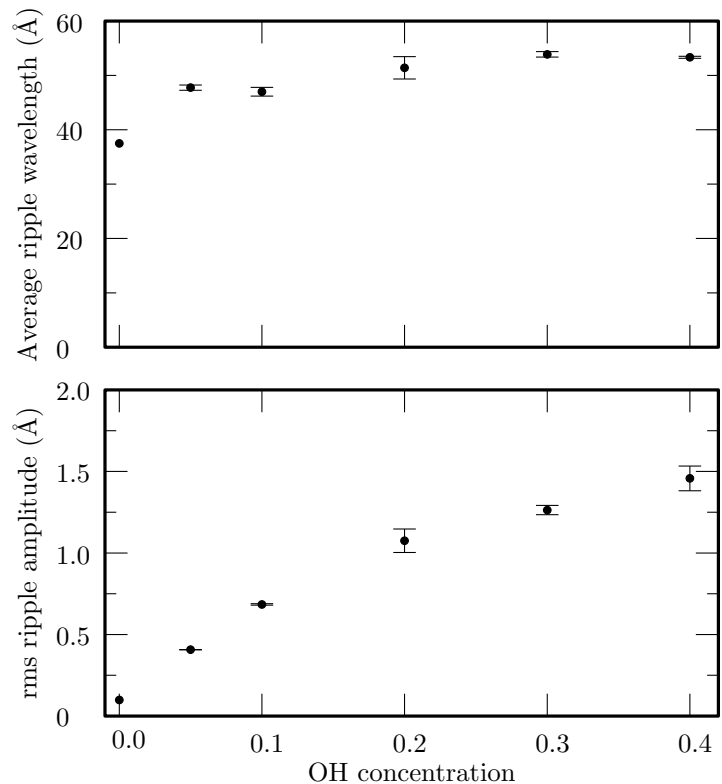}
\par\end{centering}

\caption{Wavelength and amplitude of buckles in 200$\times200$ $\mbox{\AA}$
sheets as a function of OH concentration. The wavelength changes rather
little with concentration, while rms amplitude increases. Amplitudes
of the rippled peaks are around six times larger than the rms amplitude;
peak-to-peak amplitude at 40\% OH concentration is around 10 $\mbox{\AA}$.
Wavelength and rms amplituded were computed for free-standing sheets
after excluding 20 $\mbox{\AA}$ of material at the edge of the sample.
Wavelength was computed by decomposing the sheet into a series of
line scans, taking the one-dimensional Fourier transform of them in
turn, finding the average wave vector $\bar{k}$ for each line weighted
by the amplitude of the Fourier transform, computing $\lambda=2\pi/\bar{k}$
and finally averaging $\lambda$ over all the line scans. Error bars
represent standard errors after averaging over three independent trials
per concentration. \label{fig:Wavelength-and-amplitude}}

\end{figure}

As shown in Figure \ref{fig:Ripples-in-graphene2}, with a 20\% concentration
of OH adsorbates, graphene develops ripples whose wavelength and amplitude
are comparable to those seen in experiment. Peak-to-peak amplitude
of ripples is around six times greater than the rms amplitude, so
by 40\% concentration, the ripples have a peak-to-peak amplitude of
around 10 $\mbox{\AA}$. The variation of ripple amplitude and wavelength
is displayed in Figure \ref{fig:Wavelength-and-amplitude}. It is
not possible to deduce the OH concentration from these computations
because both experimental and computational uncertainties are too
large at this point to permit it. Whether this mechanism is in fact
responsible for the buckles might be determined by experiments on
adsorbate-free surfaces conducted in high vacuum.

\begin{acknowledgments}
We thank Allan Macdonald for pointing this problem out to us, Andre
Geim for correspondence concerning the possible mechanism for buckling,
and the National Science Foundation for funding through DMR 0701373.
\end{acknowledgments}
\bibliographystyle{apsrev}
\bibliography{/home/marder/references/membrane,/home/marder/references/fracture,/home/marder/book/book}

\end{document}